# AN ALGORITHM TO SELF-EXTRACT SECONDARY KEYWORDS AND THEIR COMBINATIONS BASED ON ABSTRACTS COLLECTED USING PRIMARY KEYWORDS FROM ONLINE DIGITAL LIBRARIES

Natarajan Meghanathan[1], Nataliya Kostyuk[2], Raphael Isokpehi[2] and Hari Cohly[2]

[1]Department of Computer Science, [2]Department of Biology, Jackson State University,
1400 John Lynch St, Jackson, MS 39217, USA
natarajan.meghanathan@jsums.edu

## ABSTRACT

*The high-level contribution of this paper is the development and implementation of an algorithm to self-extract secondary keywords and their combinations (combo words) based on abstracts collected using standard primary keywords for research areas from reputed online digital libraries like IEEE Explore, PubMed Central and etc. Given a collection of N abstracts, we arbitrarily select M abstracts (M<< N; M/N as low as 0.15) and parse each of the M abstracts, word by word. Upon the first-time appearance of a word, we query the user for classifying the word into an Accept-List or non-Accept-List. The effectiveness of the training approach is evaluated by measuring the percentage of words for which the user is queried for classification when the algorithm parses through the words of each of the M abstracts. We observed that as M grows larger, the percentage of words for which the user is queried for classification reduces drastically. After the list of acceptable words is built by parsing the M abstracts, we now parse all the N abstracts, word by word, and count the frequency of appearance of each of the words in Accept-List in these N abstracts. We also construct a Combo-Accept-List comprising of all possible combinations of the single keywords in Accept-List and parse all the N abstracts, two successive words (combo word) at a time, and count the frequency of appearance of each of the combo words in the Combo-Accept-List in these N abstracts.*

## 1. INTRODUCTION

Each research area has its own set of keywords that need to be used to extract relevant documents of interest. This often requires the help of human subject matter experts who list the set of keywords that can be used for document search in a research area. However, with rapid evolution of different fields of research, it becomes inevitable to automate the process of identifying keywords based on the abstracts and publications available for specific research areas, rather than relying on availability of human experts, In this paper, we develop an algorithm to for self-extracting secondary keywords and their combinations, referred to as combo words, based on abstracts that have been collected from online digital libraries using standard primary keywords. Our proposed algorithm efficiently parses a randomly selected subset of the collection of abstracts and uses the list of acceptable single keywords, formed through the training, to count the frequency of these single secondary keywords and their combo words in all of the abstracts. The acceptable list of single secondary keywords is constructed by parsing the randomly selected abstracts, word by word, and querying the user whether to accept or not accept a word when seen for the first time. As we parse through more abstracts, we observe that the number of times the user is queried for classification reduces drastically.

The identification of most commonly used unique secondary keywords and their combo words will aid in the efficient search and filtering of the vast amount of information (like abstracts and publications) available for specific research areas in well-known online digital libraries. Eventually, our proposed approach can be used in the construction of an ontology tree of keywords for specific research areas. The rest of the paper is organized as follows: In Section 2, we describe in detail the algorithm for self-extraction of secondary keywords and their combo words. Section 3 presents the evaluation of the proposed algorithm and the results obtained for the frequency distribution of secondary keywords and their combo words in the following three research areas: Sensor Networks, Autism and Language Development, Autism and Genetics. Section 4 lists related work. Section 5 concludes the paper and also lists our future research direction.

## 2. ALGORITHM TO SELF-EXTRACT KEYWORDS

The proposed algorithm first constructs an acceptable list of secondary keywords (*Accept-List*) by parsing *M* randomly selected abstracts out of the total of *N* abstracts collected for a specific research area from online digital libraries. The algorithm operates using two datasets: a set of acceptable keywords (referred to as *Accept-List*) and a set of non-acceptable words (referred to as *Non-Accept-List*). Initially both these datasets are empty. We parse each of the *M* abstracts, word by word. If a word parsed by the algorithm is neither in the *Accept-List* nor in the *Non-Accept-List*, we ask the user to classify the word into one of these two lists. The word is appropriately appended to either of the two lists selected by the user. If the algorithm encounters a word that is already in one of the two lists, then the user is not queried again. After the set of acceptable words is built by parsing the *M* abstracts, we now parse all the *N* abstracts, word by word, and count the frequency of appearance of each of the words in *Accept-List* in these *N* abstracts. If an acceptable word appears in an abstract, we increment its frequency counter by one, irrespective of the number of times the word appears in the abstract. The algorithm outputs the secondary keywords from the *Accept-List* in the decreasing order of the frequency of appearance of the words in the *N* abstracts. The pseudo code for returning the frequency of appearance of the secondary keywords in illustrated in Figure 1. We also construct a *Combo-Accept-List* comprising of all possible combinations of the single keywords in *Accept-List* and parse all the *N* abstracts, two successive words (combo word) at a time, and count the frequency of appearance of each of the combo words in *Combo-Accept-List* in these *N* abstracts. The pseudo code for returning the frequency of appearance of the secondary combo keywords in illustrated in Figure 2.

---

**Input:** Set *A* – set of abstracts in text format; total number of abstracts $N = |A|$
 *M* – size of the training set of abstracts
 *Separator-List* = { , . ; - ' < > ( ) [ ] / " : }
**Auxiliary Variables:** Training-Set *TS* – set of *M* abstracts randomly selected from *A*; $TS \subseteq A$
 *Accept-List* – set of keywords acceptable as secondary keywords
 *Non-Accept-List* – set of words not acceptable as secondary keywords
 *User-Choice* = {0 or 1} for a word
 *Already-Accounted-Words* – set of keywords in *Accept-List* already accounted for their presence in a particular abstract
**Output:** Set *Freq-Sec-KW* – set storing the frequency of appearance of each word in the *Accept-List* in the decreasing order of the frequency of their appearance in the *N* abstracts

**Initialization:** *Accept-List* ← Φ
               *Non-Accept-List* ←Φ
               Training-Set *TS* ← Φ

**Begin** *Self-Extract-Key-Words*

// Form the training set of abstracts
1   **while** (|*TS*| < *M*) **do**
2      Generate a random integer *m*∈ [1…*N*]
3      **if** *A*[*m*]∉ *TS* **then**
4        *TS* ← *TS* U *A*[*m*]
5      **end if**
6   **end while**

// Build the *Accept-List* and *Non-Accept-List*
7   **for** every abstract *P* ∈ *TS* **do**
8     **for** every word *W*∈ *P* **do**
9       **if** *W* ∉ *Separator-List* **then**
10        **if** *W*∉ *Accept-List* AND *W*∉ *Non-Accept-List* **then**
11          *User-choice* ← Ask user to enter 0 to put the word in *Non-Accept-List* or 1 for
                  *Accept-List*
12          **if** *User-choice* = 1 **then**
13            *Accept-List* ← *Accept-List* U {*W*}
14          **else**
15            *Non-Accept-List* ← *Non-Accept-List* U {*W*}
16          **end if**
17        **end if**
18      **end if**
19     **end for**
20   **end for**

// Count the frequency of appearance of the words in *Accept-List*
21   **for** every word *W*∈ *Accept-List*
22     *Freq-Sec-KW*(*W*) ←0
23   **end for**
24   **for** every abstract *Q*∈*A* **do**
25      *Already-Accounted-Words* ← Φ
26      **for** every word *W*∈*Q* **do**
27       **if** *W*∉ *Separator-List* AND *W*∉ *Non-Accept-List* **then**
28        **if** *W*∉ *Already-Accounted-Words* AND *W*∈ *Accept-List* **then**
29          *Freq-Sec-KW*(*W*) ← *Freq-Sec-KW*(*W*) + 1
30          *Already-Accounted-Words* ← *Already-Accounted-Words* U {*W*}
31        **end if**
32       **end if**
33      **end for**
34   **end for**

35   **Sort** the words in *Freq-Sec-KW* in the decreasing order of the frequency of appearance of
       the secondary keywords

**return** *Freq-Sec-KW*

**End** *Self-Extract-Key-Words*

---

**Figure 1:** Algorithm to Self-Extract Single Secondary Keywords

---

**Input:** *Accept-List* // list of acceptable keywords
       Set *A* – set of abstracts in text format; total number of abstracts *N* = |*A*|
       *Separator-List* = { , . ; - ' < > ( ) [ ] / " : }
**Auxiliary Variables:** *Already-Accounted-Combo-Words* – set of keywords in *Accept-List*
                already accounted for their presence in a particular abstract
                *Combo-Accept-List* // list of acceptable combinations of keywords
**Output:** Set *Freq-Comb-Sec-KW* // – set storing the frequency of appearance of combination of
       any two keywords in the *Accept-List* in the decreasing order of the frequency of their
       appearance in the *N* abstracts

**Initialization:** *Combo-Accept-List* ← Φ

**Begin** *Self-Extract-Combo-Key-Words*

// Build the *Combo-Accept-List*
    1  **for** every index *i* ∈ [0…|*Accept-List*|-1] **do**
    2   **for** every index *j* ∈ [i+1….|*Accept-List*|] **do**
    3     W1 ← *Accept-List*[*i*]
    4     W2 ← *Accept-List*[*j*]
    5     W12 = W1 W2
    6     W21 = W2 W1
    7     *Combo-Accept-List* = *Combo-Accept-List* U {W12}
    8     *Combo-Accept-List* = *Combo-Accept-List* U {W21}
    9   **end for**
  10  **end for**

// Count the frequency of appearance of combination of keywords in *Combo-Accept-List*
  11  **for** every combo word $W_{ij}$ ∈ *Combo-Accept-List*
  12    *Freq-Combo-Sec-KW*($W_{ij}$) ← 0
  13  **end for**
  14  **for** every abstract *Q* ∈ *A* **do**
  15  *Already-Accounted-Combo-Words* ← Φ
  16   **for** every combo word *W* ∈ *Q* **do**
  17    **if** *W* ∉ *Separator-List* AND *W* ∈ *Combo-Accept-List* **then**
  18     **if** *W* ∉ *Already-Accounted-Combo-Words* **then**
  19      *Freq-Combo-Sec-KW*(*W*) ← *Freq-Combo-Sec-KW*(*W*) + 1
  20      *Already-Accounted-Combo-Words* ← *Already-Accounted-Combo-Words* U {*W*}
  21     **end if**
  22    **end if**
  23   **end for**
  24  **end for**

  25  **Sort** the words in *Freq-Combo-Sec-KW* in the decreasing order of the frequency of
       appearance of the combination of secondary keywords

**return** *Freq-Combo-Sec-KW*

**End** *Self-Extract-Combo-Key-Words*

**Figure 2:** Algorithm to Self-Extract Combo Words (Combinations of Secondary Keywords)

## 3. ALGORITHM EVALUATION AND RESULTS

We implemented the proposed algorithm in Java. We selected the following three research areas for testing our algorithm: (i) Sensor Networks – SN, (ii) Autism and Language Development – ALD and (iii) Autism and Genetics – AG. We used the well-known digital libraries like IEEE Explore (for Sensor Networks) [1] and PubMed (for Autism and Language Development; Autism and Genetics) [2] for collecting the abstracts. For each of these research areas, we collected $N = 100$ abstracts and constructed the *Accept-List* by letting $M = 15$, 30 and 50 abstracts. After the construction of the *Accept-List* for a particular value of $M$, we parsed all the $N$ abstracts and counted the frequency of appearance of each of the keywords in *Accept-List*, in each of the $N$ abstracts. Also, for each value of $M$, we constructed a *Combo-Accept-List* comprising of all possible combinations of the single keywords in *Accept-List* and parsed all the $N$ abstracts, two successive words (combo word) at a time, and counted the frequency of appearance of each of the combo words in the *Combo-Accept-List* in these $N$ abstracts. While computing the frequency distribution, a keyword in the *Accept-List* is counted only once for an abstract, even though the keyword may appear more than once. The same rule is applied while computing the frequency of the combo words. In other words, the frequency of appearance of a keyword and a combo word can be at most $N$, the total number of abstracts parsed.

### 3.1 Effectiveness of the Training Approach

The effectiveness of our training approach is measured by measuring the percentage of words for which the user is queried for classification when the algorithm parses through the words of each of the $M$ abstracts. For each of the three research areas, we observed that as $M$ grows larger, the percentage of words for which the user is queried for classification reduces drastically. This is illustrated through Figures 3.1, 3.2 and 3.3. We notice in these three figures that the slope of the trend line for all the three sets of abstracts decreases drastically as $M$ grows from small to moderate values. As M grows from moderate to larger values, the trend line becomes flat (i.e., slope equals 0), indicating that smaller to moderate values of $M$ is sufficient to construct the *Accept-List*.

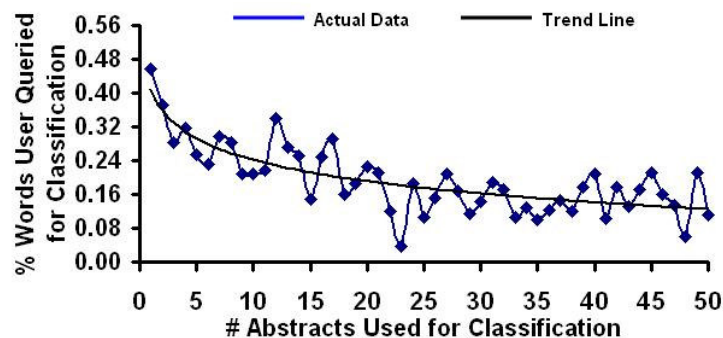

**Fig 3.1:** Results for Sensor Networks

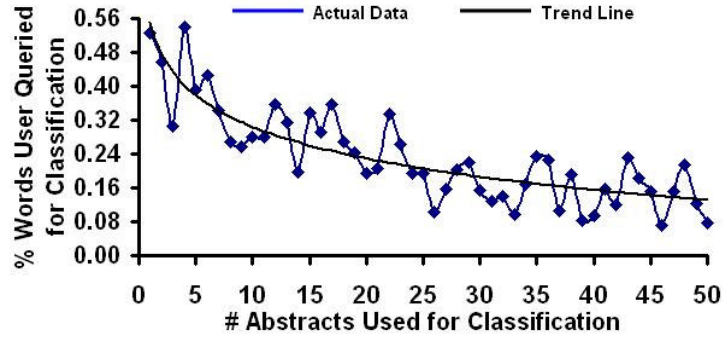

**Fig 3.2:** Results for Autism and Language Development

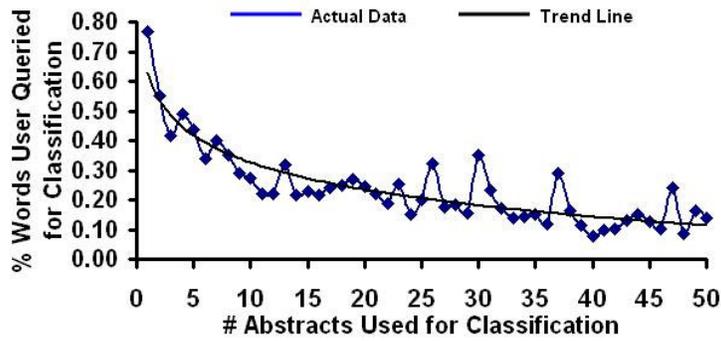

**Fig 3.3:** Results for Autism and Genetics

**Figure 3:** Percentage of Words the Self-Extraction Algorithm Queries the User for Classification Vs the Number of Abstracts Used for Classification

**Table 1:** Top 15 Single Keywords and Top 15 Combo Words of Single Keywords Extracted from IEEE Abstracts with the Search String: "Wireless Sensor Networks"

| # | Top 15 Single Keywords | Freq | Top 15 Combo Words of Single Keywords | Freq |
|---|---|---|---|---|
| 1 | Node | 65 | Routing Protocol | 16 |
| 2 | Wireless | 64 | Ad hoc | 15 |
| 3 | Application | 59 | Wireless Communication | 8 |
| 4 | Base | 48 | Energy Consumption | 7 |
| 5 | Design | 43 | Environmental Monitoring | 7 |
| 6 | Protocol | 41 | Power Consumption | 6 |
| 7 | Data | 40 | Sensing Computation | 5 |
| 8 | Energy | 38 | Real time | 5 |
| 9 | Communication | 38 | Malicious Attack | 5 |
| 10 | Deploy | 37 | Battery Power | 5 |
| 11 | System | 36 | Sink Node | 4 |
| 12 | Performance | 35 | Layer Protocol | 4 |
| 13 | Simulation | 35 | MAC Layer | 4 |
| 14 | Computation | 32 | Data Fusion | 4 |
| 15 | Time | 30 | Topology Control | 4 |

**3.2 Frequency of Appearance of the Secondary Keywords and their Combo Words**

We noticed that the frequency of appearance of the top 15 keywords and the top 15 combo words of these secondary keywords in the N=100 abstracts for a specific research area remains the same for different values of *M* with a confidence interval above 95%. Hence, the data presented in Tables 1, 2 and 3 illustrate the frequency of appearance of the secondary keywords and their combo words obtained for *M* = 15. The results are the same for *M* = 30 and *M* = 50 randomly selected abstracts.

**Table 2:** Top 15 Single Keywords and Top 15 Combo Words of Single Keywords Extracted from PubMed Abstracts with the Search String: "Autism and Language Development"

| # | Top 15 Single Keywords | Freq | Top 15 Combo Words of Single Keywords | Freq |
|---|---|---|---|---|
| 1 | Disorder | 69 | Autism Spectrum Disorder | 49 |
| 2 | Children | 69 | Development Disorder | 14 |
| 3 | Spectrum | 51 | Social Interaction | 14 |
| 4 | Age | 46 | Young Child | 13 |
| 5 | Behaviour | 43 | Language Impairment | 11 |
| 6 | Social | 43 | Development Delay | 11 |
| 7 | ASD | 39 | Pervasive Development | 9 |
| 8 | Diagnosis | 35 | PDD nos | 7 |
| 9 | Assessment | 32 | Impairment SLI | 7 |
| 10 | Communication | 31 | Receptive Language | 7 |
| 11 | Function | 31 | ASD Group | 7 |
| 12 | Cognition | 30 | Language Ability | 7 |
| 13 | Association | 29 | Language Disorder | 7 |
| 14 | Clinical | 29 | Verbal IQ | 6 |
| 15 | Delay | 27 | Social Behaviour | 6 |

**Table 3:** Top 15 Single Keywords and Top 15 Combo Words of Single Keywords Extracted from PubMed Abstracts with the Search String: "Autism and Genetics"

| # | Top 15 Single Keywords | Freq | Top 15 Combo Words of Single Keywords | Freq |
|---|---|---|---|---|
| 1 | Disorder | 75 | Autism Spectrum Disorder | 32 |
| 2 | Association | 53 | Mental Retardation | 20 |
| 3 | Phenotype | 37 | Neurodevelopment Disorder | 13 |
| 4 | Spectrum | 36 | X Syndrome | 9 |
| 5 | Syndrome | 33 | Nucleotide Polymorphism | 9 |
| 6 | Patient | 31 | Bipolar Disorder | 8 |
| 7 | Family | 31 | Risk Factor | 6 |
| 8 | Behavior | 30 | Hyperactivity Disorder | 6 |
| 9 | Neuron | 29 | Gene Expression | 6 |
| 10 | Genome | 28 | Binding Protein | 5 |
| 11 | Clinical | 28 | Gene Factor | 5 |
| 12 | Cause | 28 | Molecular Gene | 5 |
| 13 | Control | 26 | Deficiency Hyperactivity | 5 |
| 14 | Mental | 25 | Psychiatric Disorder | 5 |
| 15 | Chromosome | 24 | Chromosome Abnormality | 5 |

## 4. RELATED WORK

In [3], a model for extracting keywords from abstracts and titles was proposed and tested on a set of abstracts of academic papers containing keywords composed by their authors. However, this model does not adopt an effective training module to build a list of acceptable keywords and lacks the ability to prioritize the keywords based on their frequency of appearance in the abstracts. In [4], the authors presented a model for automatic annotation of protein functions based on biological information directly extracted from MEDLINE [5] abstracts. However, unlike our proposed algorithm in this paper, the model proposed in [4] is not generic and has been developed to specifically extract domain-specific information from the analysis of a set of protein families.

Usui et. al [6] developed a tool to customize the index tree of different Neuroinformatics platforms by allowing a user to recognize the most relevant terms and easily incorporate the preferred terms over the list of suggested terms. However, this model is too memory consuming as it requires the use of a large two-dimensional matrix whose row index represents the keywords, filtered after stop word removal and stemming, column index corresponds to the documents and the entries in the matrix correspond to the number of times each keyword appears in a document. Also, the list of keywords returned by their model is hypothetical and heavily depends on the weighting scheme used to prioritize the entries in the matrix. On the other hand, our proposed algorithm uses a simple single-dimension list, *Accept-List*, of keywords formed through linear parsing of the abstracts. Also, we exclusively rely on counting the frequency of appearance of the acceptable list of secondary keywords in the entire set of abstracts collected for a specific research area.

## 5. CONCLUSIONS AND FUTURE RESEARCH DIRECTION

The proposed algorithm can be used to efficiently extract secondary keywords based on abstracts collected using the names of the specific research areas as the primary keywords from online digital libraries. The algorithm works with abstracts as text files and parses the abstracts word by word using pre-defined separator lists. The effectiveness of our training approach to build the acceptable list of secondary keywords is vindicated by the exponential decreasing trend of the percentage number of words for which the user is queried for classification, as we increase the number of abstracts used for training. For each of the three research areas explored in this paper, we observed that a mere 15% of abstracts randomly selected from the set of all abstracts are sufficient to effectively identify and rank the secondary keywords and their combo words. The algorithm builds the acceptable list on a randomly generated subset of the collection of abstracts and hence the secondary keywords extracted are representative of the entire research area. The top 15 secondary keywords and their combo words listed in Tables 1, 2 and 3 are widely used terminologies and words in the specific research areas. The application of the algorithm has been tested for two widely different domains of research: Sensor Networks and Autism.

In the near future, we plan to use the algorithm to construct an ontology tree and directed acyclic graph of keywords for different research areas and this can be used for efficient literature search and effective retrieval of documents from online digital libraries that store vast amount of information. We will adapt the pseudo code in Figure 1 to form an acceptable list of combo words, directly extracted from a random subset of the abstracts, and then determine the frequency distribution of these combo words. We will compare such a frequency distribution of combo words with that obtained for combo words based on combinations of secondary keywords.

## 6. ACKNOWLEDGMENTS

Mississippi NSF-EPSCoR (EPS-0556308, EPS-0903787); Pittsburgh Supercomputing Center's National Resource for Biomedical Supercomputing (T36 GM008789); U.S. Department of Homeland Security Science & Technology Directorate (2007-ST-104-000007, 2009-ST-062-000014, 2009-ST-104-000021); Mississippi Computational Biology Seed Research Grant Program; and Research Centers in Minority Institutions (RCMI) – Center for Environmental Health (NIH-NCRR G12RR13459). The views and conclusions contained in this document are those of the authors and should not be interpreted as necessarily representing the official policies, either expressed or implied, of the funding agency.